\documentclass[onecolumn,pre] {revtex4}
\usepackage{graphics}
\usepackage{epsf}
\usepackage{epsfig}
\usepackage{amsmath}
\usepackage{amssymb}
\usepackage{amscd}

\begin{document}

\newcommand{\ACAL}{\mathcal{A}} %  mathcal
\newcommand{\fig}[1]{Fig.~\ref{f:#1}}
\newcommand{\eq}[1]{eq.~(\ref{e:#1})}
\newcommand{\sma}{\text{s}}
\newcommand{\lar}{\text{l}}
\newcommand{\eff}{\text{eff}}
\newcommand{\refer}{\text{ref}}
\newcommand{\bulk}{\text{b}}
\title{Selective-pivot sampling of radial distribution functions in
 asymmetric liquid mixtures}
\author{J. G. Malherbe} 
\affiliation{Physique des Liquides et Milieux Complexes, Facult\'{e}
des Sciences et de Technologie,Universit\'{e} Paris XII, 61 av. du
G\'{e}n\'{e}ral de Gaulle, 94010 Cr\'{e}teil Cedex, France}
\author{Werner Krauth}
\affiliation{CNRS-Laboratoire de Physique Statistique, Ecole Normale
Sup\'{e}rieure, 24 rue Lhomond, 75231 Paris Cedex 05, France}

\date{29 May 2007}

\begin{abstract}
We present a Monte Carlo algorithm for selectively sampling radial
distribution functions and effective interaction potentials in asymmetric
liquid mixtures.  We demonstrate its efficiency for hard-sphere mixtures,
and for model systems with more general interactions, and compare our
simulations with several analytical approximations. For interaction
potentials containing a hard-sphere contribution, the algorithm yields
the contact value of the radial distribution function.  
\end{abstract}

\maketitle

\section{Introduction}
Liquid mixtures have been studied in the past decades with objectives
ranging from understanding basic theoretical properties to answering
questions of technological relevance. The simplest model of liquid
mixtures, binary hard spheres, has been an important test bed for
experimental, analytical and numerical techniques.

\begin{figure}[htpb]
\centerline{
\epsfxsize=8.0cm
\epsfclipon
\epsfbox{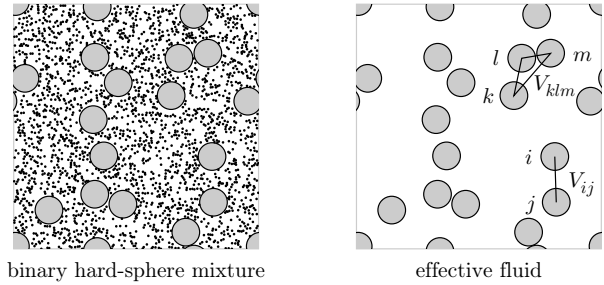}}
\caption{Asymmetric hard-sphere mixture 
(\emph{left}), and its
effective fluid of solutes (pair interaction $V_{ij}$ and triplet interaction $V_{klm}$ shown,
 \emph{right}).}
\label{f:schema_effective}
\end{figure}

In asymmetric binary mixtures consisting of few large particles (solutes)
and many small particles (solvent) (see \fig{schema_effective}), one may
in principle integrate out all the degrees of freedom of the solvent,
and arrive at an equivalent effective fluid of only the large particles,
with complicated effective interactions involving pairs, triplets,
and, more generally, $n$-tuples of solutes \cite{Hill}. Usually,
all contributions beyond the pair interaction are neglected (see,
for example, \cite{Dijkstra}), even though they are quite important
\cite{Tehver,JGMP2,SJPC2}. Moreover, integral-equation theories or density
functional theories \cite{EvansDFT}, among others, allow to study the
mixtures directly.  For asymmetric hard-sphere mixtures, various approaches
based on thermodynamic criteria yield 
empirical equations of state 
(see for example \cite{gij-cont}).  However, many theories may fail
in extended regions of parameter space because of the different sizes of
solutes and solvent (for problems with the integral-equation theories,
see \cite{Ayadim}).

Numerical simulation is a crucial tool to validate the above methods.
However it  also meets difficulties for asymmetric mixtures.  On the
one hand, the displacement of each solute is highly constrained
by the surrounding solvent (see \cite{SMAC}). Pivot-cluster Monte
Carlo algorithms \cite{Dress} have partially overcome this problem
for hard-sphere models \cite{Krauth} and for interacting mixtures
\cite{JGMP1,Liu1}, as long as the overall density of the system is
not too high \cite{Liu2}.  On the other hand, the difference in size
of solutes and of solvent particles generates a sampling problem for
observables such as the pair-distribution functions, which vary strongly
on the very small scale of the solvent. To describe these observables,
one needs many data  (generally contained in a histogram with a 
fine grid). On close approach of two solutes in asymmetric mixtures,
the pair correlation functions rise steeply. Therefore, the contact
value of the pair-correlation function cannot be obtained precisely
by extrapolation \cite{Revue Allen}.

In this paper, we introduce a selective-pivot sampling algorithm, which
allows us to compute the ratio of the radial distribution functions
$g(R)/g(R')$ for two arbitrary inter-particle distances $R$ and $R'$.
For hard-sphere mixtures, we compute the pair-correlation function of
the solutes and especially its contact value. We also determine the
potential of the mean force. We furthermore discuss the extension of
the method to systems with arbitrary pair interactions. 

\section{ Selective-pivot sampling}

\begin{figure}[htbp]
\centerline{
\epsfclipon
\epsfbox{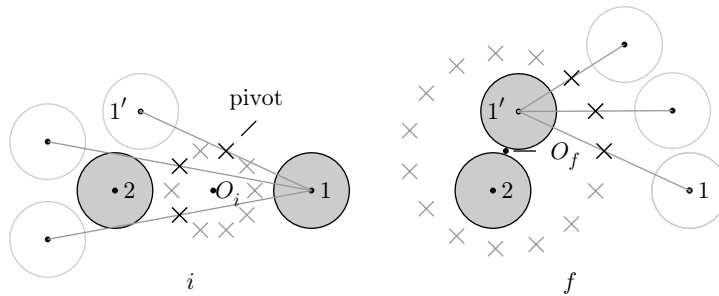}}
\caption{
   Selective-pivot move $1 \to 1'$ (configuration $i$, \emph{left}) and
   return move $1' \to 1$ ($f$, \emph{right}).  In $i$,
   the distance $R_{12}= R_i$, and $R_{1'2}= R_f$. All
   pivots resulting in a new distance $R_f$ lie on a sphere $S_i$
   with center $O_i$ and radius $R_f/2$ 
   (\emph{crosses}).}
   \label{f:selective_pivot}
\end{figure}

For simplicity, we first discuss selective-pivot sampling for a system
of an isolated  pair of three-dimensional hard spheres. To sample the ratio
 of the pair
correlation function for two arbitrarily distances, $R_i$ and $R_f$,
we consider a configuration $i$, with a distance $R_i$ between the two
particles, and all moves of one particle ($1 \to 1'$) such that the
new distance is $R_f$ (see \fig{selective_pivot}). Any such move can be
interpreted as a reflection around a pivot located on a sphere $S_i$ with
radius $R_f/2$ and center $O_i$, half-way between the two particles. We
choose the pivot randomly on the sphere $S_i$ (using Gaussian random
numbers, see \cite{SMAC}). The a priori probabilities for proposing the
move $i \to f$, and the return move $f \to i$ are, respectively,
\begin{equation*}
\ACAL(i \to f) = \frac{1}{S_i} \propto \frac{1}{R_f^2}, \quad
\ACAL(f \to i) = \frac{1}{S_f} \propto \frac{1}{R_i^2}.
\end{equation*}
The detailed-balance condition connects the stationary probabilities $\pi(i)$ and 
$\pi(f)$ with the a priori probabilities:
\begin{equation*}
\pi(i)  \ACAL(i \to f)= \pi(f)  \ACAL(f \to i).
\end{equation*}
It follows that the probabilities of the two configurations are biased
by a geometric factor with respect to the constant hard-sphere
probabilities,
\begin{equation}
\frac{\pi(i)}{\pi(f)} = \frac{S_f}{S_i} = \frac{R_f^2}{R_i^2}. 
\label{e:ratio_of_probabilities}
\end{equation}
For spherically symmetric systems, the pair correlation function
depends solely on the distance between particles and reduces to the 
radial distribution function $g(R)$, which 
is linked with the probability $\pi(R)$ to observe a particle
in an infinitesimal spherical shell of radius $R$, the
other one being at its center, by 
\begin{equation*}
\pi(R) \propto R^{2} g(R)\text{d}R.
\end{equation*}
The biasing factor in \eq{ratio_of_probabilities} 
cancels the phase-space factor of the radial distribution function, 
and the ratio of the probabilities $\pi(i)$ and $\pi(f)$ equals the 
ratio of the radial distribution functions: 
\begin{equation}
\frac{\pi(i)}{\pi(f)} = \frac{g_i}{g_f}.
\label{e:probabilities_radial}
\end{equation}
The above algorithm and the relation of \eq{probabilities_radial}
between the probabilities of observing the distances $R_i$ and $R_f$
remain valid for two or more solute particles in a box with periodic
boundary conditions, in the presence of other components (solvent). It
suffices to permanently tag two solutes.  One Monte Carlo
move involves a single pivot, but may transform many particles (see
\cite{SMAC}), including both tagged ones, so that the distance between
them may  not change.

\section{Applications}

To validate the selective-pivot sampling algorithm, and to explore
possible applications, we compute the distribution function of the
solutes for asymmetric binary mixtures, and compare it with simulation
data obtained with standard cluster simulation methods, as well as with
analytic approximations. In this system, we also compute the contact
value of the distribution function. This can be done directly, without
extrapolation. Furthermore, we use the selective-pivot sampling algorithm
to compute the potential of the mean force, that is, the distribution
function for two solutes in a large bath of solvent, approaching the
regime of infinite dilution.

\subsection{Radial distribution function of solutes}

We consider an asymmetric binary mixture of spheres with
diameters $D_{\sma}$ (small particles---solvent) and $D_{\lar}$
(large particles--solute), and size ratio $R=D_\lar/D_\sma=10$,
at packing fractions $\eta_\sma=0.126$ and $\eta_\lar=0.121$, where
the cluster algorithm  performs well (the packing fraction of the
solvent, $\eta_\sma$, is linked to the number of solvents $N_\sma$
in the simulation box of volume $V $ by $\eta_\sma=\frac{\pi}{6}
(N_\sma/V) D_\sma^3$, \emph{etc.}).

\begin{figure}[htbp]
\centerline{
\epsfxsize=7.0cm
\epsfclipon
\epsfbox{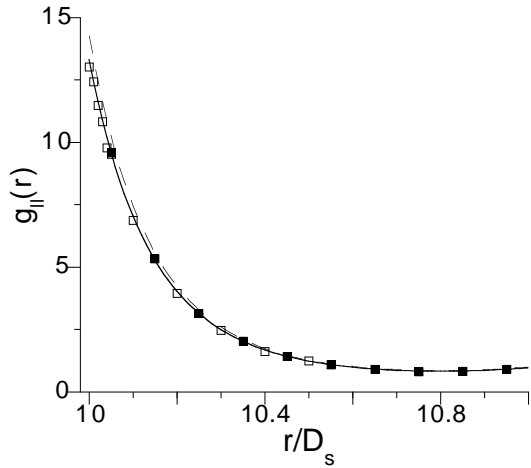}}
\caption{Radial distribution function of the solutes for hard spheres 
with $R=10$,
$\eta_\sma=0.126$ and $\eta_\lar=0.121$. \emph{Symbols}: Cluster algorithm
 (\emph{filled squares}:
standard sampling method; \emph{empty squares}: selective-pivot sampling); 
\emph{full line}:
density functional theory; \emph{dashed line}: RHNC-MSA \cite{Ayadim}}
\label{f:10mel}
\end{figure}
In \fig{10mel}, we compare the radial distribution function of the large
spheres $g_{\lar \lar}(r)$ obtained from selective-pivot sampling and from
a standard cluster algorithm and find very good agreement.  The contact
region and, especially, the contact value $g_{\lar \lar}(r=D_\lar)$ become
accessible only with our sampling scheme.  With standard grid sampling,
they can be obtained only by extrapolation because the probability
of observing two solutes exactly at contact is zero.  Our sampling
approach gives sensibly more accurate contact values than the standard
method (see \cite{grcont}). In \fig{10mel}, $g_{\lar \lar}(D_\lar)=13.03 
\pm 0.26$. As a common use of simulation
data, we also compare the
pair-correlation function with density functional theory, which agrees
very well \cite{YuWu,Roth-CS}, and with the Ornstein-Zernike integral
equation with the RHNC-MSA closure, which overshoots in the contact region
\cite{Ayadim}.

\subsection{ Potential of the mean force}
As another application of selective-pivot sampling, 
we consider the potential of the mean force, the solute
pair-correlation function in the limit of infinite dilution of the solute
 particles:
\begin{equation}
\lim\limits_{\eta _{\lar}\rightarrow 0}g_{\lar \lar}\left( r,\eta^{\bulk}\right)
 =\exp\left[-\beta U(r) \label{g22/pot} \right], 
\end{equation} 
where $U(r)=u_{\lar \lar}(r)+\Phi_{\eff}\left(r\right)$ is the total pair 
interaction
between the solutes, $u_{\lar \lar}(r)$ the direct interaction,
$\Phi_{\eff}(r)$ the pair potential of the mean force and $\eta_{b}$ the
bulk solvent packing fraction.  The selective-pivot sampling algorithm
allows us therefore to compute the potential of the mean force of the
effective fluid obtained by integrating out the solvent particles of the
binary mixture. One simply puts the two tagged solutes into a bath of
solvent and determines the potential differences for several pairs $R_i,
R_f$. The size of the bath plays almost no role in this simulation because
the two large particles are kept at two fixed distances, and cannot
escape to infinite separation.  The simulation can thus take place in a
large box, approaching the infinite dilution of the two solute particles.

\begin{figure}[htbp]
\centerline{
\epsfxsize=7.0cm
\epsfclipon
\epsfbox{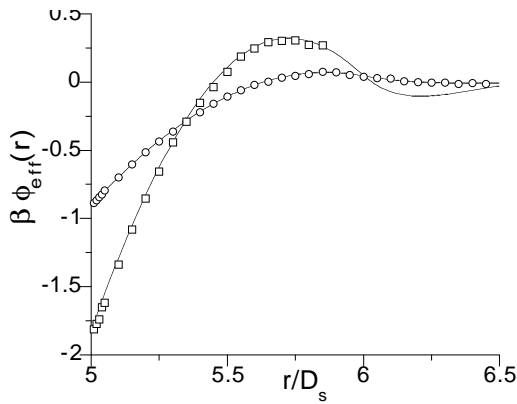}}
\caption{Potential of the mean force for hard-sphere mixture with $R=5$; 
\emph{symbols}: Selective-pivot sampling; \emph{lines}: RHNC/fundamental measure functional;
\emph{circles}: bulk packing fraction $\eta_\bulk=0.1$;
 \emph{squares}: $\eta_\bulk=0.2$. }
\label{f:fig.poteff}
\end{figure}

To illustrate this point, we computed the potential of the mean force
between two hard-sphere solutes in a bath of
hard-sphere solvent with size ration $R=5$ and bulk packing fractions
$\eta_\bulk=0.1$ and $\eta_\bulk=0.2$. We considered here $N_\sma=5280$ and  
$N_\sma=10560$ 
respectively in a cubic box of $L=30 D_\sma$.
At a difference of other simulation methods for computing the effective
interaction, this method
needs no extrapolation \cite{Dickman}  and has
no adjustable parameter \cite{Ma}.  The Monte Carlo data agree very well
with the potential of the mean force obtained from the Ornstein-Zernike
equation with the Rosenfeld fundamental measure closure \cite{Rosenf1,SJPC1}
(see \fig{fig.poteff}).

\section{Selective-pivot sampling for models with general pair interactions}

Our sampling method can be directly generalized to more general
interactions, which can also be handled by the pivot-cluster algorithm
(see \cite{Liu1,JGMP1}).  As an  illustration we study the effect of
solvation forces in a binary mixture of colloids with size ratio $R=5$,
and a Yukawa tail for unlike pairs of colloids added to the hard-sphere
interaction:
\begin{equation}
u_{\lar \sma}(r)=\left\{
\begin{array}{ll}
\infty & r<D_{\lar \sma} \\
-\epsilon _{\lar \sma}\exp \{-z_{\lar \sma}(r-D_{\lar \sma})\}/r & r
\geq D_{\lar \sma}
\end{array}
\right. . \label{yuk1}
\end{equation} 

\begin{figure}[htbp]
\centerline{
\epsfxsize=7.0cm
\epsfclipon
\epsfbox{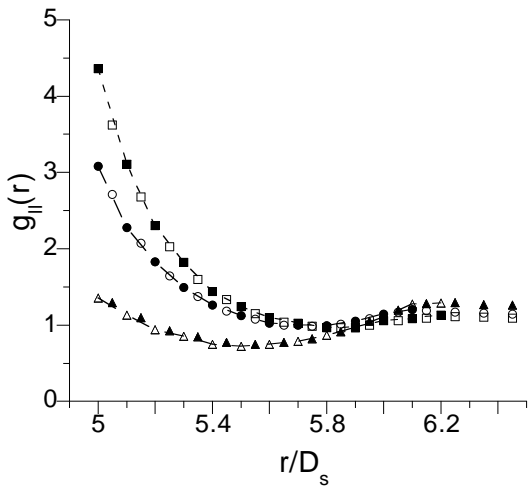}}
\caption{Pair distribution function of solute particles for $R=5$, 
$\eta_\sma=0.11$
 and $\eta_\lar=0.12$.
\emph{Empty symbols}: standard sampling method; \emph{filled symbols}: selective-pivot sampling; 
\emph{squares}: hard-sphere; \emph{circles}: Yukawa interaction with $z_{\lar \sma}^*=10$; 
\emph{triangles}: with $z_{\lar \sma}^*=2.5$.}
\label{f:fig.interac}
\end{figure}
We used a value of $0.5 k_{\text{B}} T$ for the depth of the attractive well, 
which correspond to $\epsilon _{\lar \sma}^*=\epsilon
_{\lar \sma} D_\sma/( k_{\text{B}} T)=3/2$, and considered two values for the
inverse range of the attraction force,  $z_{\lar \sma}^*=z_{\lar
\sma }D_\sma=10$ and 
$z_{\lar \sma}^*=2.5$. In \fig{fig.interac} we see that data
obtained from selective-pivot sampling and from standard cluster algorithm
agree very well.  The advantages and the applications mentioned in the
previous section carry over. \fig{fig.interac} shows that the contact
value of the radial distribution function decreases when the range of
the attraction between small and large particles increases, because 
the solvation forces create a
thin shell of small colloids which counterbalances the depletion between
the large ones.

\section{Conclusion}

In this paper, we have presented a selective sampling  cluster algorithm
which allows to obtain the radial distribution function, including the
contact value, and also yields the pair potential of the mean force
between two solutes due to the presence of the solvent. The algorithm
solves a sampling problem for observables with very rich structure on a
small scale, but does not overcome the remaining limitation of cluster
algorithms for liquid simulations, namely the restriction to moderate
densities.  It will be interesting to see whether this restriction can
also be overcome in the future.

\end{document}